\documentclass[structabstract]{aa}
\usepackage[utf8]{inputenc}
\usepackage[T1]{fontenc}
\usepackage{lmodern}
\usepackage{graphicx}
\usepackage{verbatim}
\usepackage{textcomp}
\usepackage{epsfig}
\usepackage{natbib}
\usepackage{amssymb,amsmath}

\usepackage{soul}
\usepackage{color}

\bibpunct{(}{)}{;}{a}{}{,}

\newcommand{\FIG}[1]{#1}

\begin{document}

\title{Relativistic 3D precessing jet simulations for the X-ray binary SS433}
\author{Remi Monceau-Baroux \inst{1} \and  Oliver Porth \inst{1,3} \and Zakaria Meliani \inst{2} \and Rony Keppens \inst{1}}
\institute{Centre for mathematical Plasma Astrophysics, Department of Mathematics, KU Leuven, \hfill \\
Celestijnenlaan 200B, 3001 Heverlee, Belgium \and LUTh, Observatoire de Paris, France \and Department of Applied Mathematics, The University of Leeds, Leeds, LS2 9JT}


\abstract
{Modern high-resolution radio observations allow us a closer look into the objects that power relativistic jets. This is especially the case for SS433, an X-ray binary that emits a precessing jet that is observed down to the subparsec scale.}
{We aim to study full 3D dynamics of relativistic jets associated with active aalactic nuclei or X-ray binaries (XRB). In particular, we incorporate the precessing motion of a jet into a model for the jet associated with the XRB SS433. 
Our study of the jet dynamics in this system focuses on the subparsec scales. We investigate the impact of jet precession and the variation of the Lorentz factor of the injected matter on the general 3D jet dynamics and its energy transfer to the surrounding medium. After visualizing and quantifying jet dynamics, we aim to realize synthetic radio mapping of the data, to compare our results with observations.}
{For our study we used a block-tree adaptive mesh refinement scheme and an inner time-dependent boundary prescription to inject precessing bipolar supersonic jets. Parameters extracted from observations were used. Different 3D jet realizations that match the kinetic flux of the SS433 jet were intercompared, which vary in density contrast and jet beam velocity. We tracked the energy content deposited in different regions of the domain affected by the jet. Our code allows us to follow the adiabatic cooling of a population of relativistic particles injected by the jet.
This evolving energy spectrum of accelerated electrons, using a pressure-based proxy for the magnetic field, allowed us to obtain the radio emission from our simulation.}
{We find a higher energy transfer for a precessing jet than for standing jets with otherwise identical parameters as a result of the effectively increased interaction area. We obtain synthetic radio maps for all jets, from which one can see that dynamical flow features are clearly linked with enhanced emission sites.}
{The synthetic radio map best matches with a jet model using the canonical propagation speed of $0.26c$ and a precession angle of $20^\circ$. Overdense precessing jets experience significant deceleration in their propagation through the interstellar medium (ISM), while the overall jet is of helical shape. Our results show that the kinematic model for SS433 has to be corrected for deceleration assuming ballistic propagation on a subparsec scale.}

\maketitle

\section{Introduction}

Non-relativistic up to highly relativistic jets are associated with a wide range of astrophysical objects. They can manifest themselves as a young stellar jet flows, or appear as (electromagnetically driven) jets in pulsar wind nebulae. High-luminosity jets are detected through radio observations  from active galactic nuclei (AGN) and from X-ray binaries (XRB), the latter also denoted as microquasar systems. A review of these jets and their different origins and similarities can be found in \citet{livio2000}.
The studies of AGN and XRB jets are related and are of utmost interest because their high kinetic influx is an efficient way to transfer energy from small structures to larger ones in their surroundings. Both AGN \citep[see][]{silk1998} and XRB \citep[see][]{fragos2013} jets play a role for cosmological evolutions by feeding back part of the accreted energy to the galaxy clusters.
Using numerical simulations, we aim to quantify precisely this energy transfer from the jet to its surroundings. In \citet{monceau2012}, we studied the impact that varying the jet opening angle has on the resulting jet dynamics. We here study a jet of finite opening angle with a more dramatic temporal change: a precessional motion of the jet. This implies dropping the assumption of axisymmetry, which is commonly used in various studies on relativistic jet propagation, and requires full 3D simulations. Such a precessing motion can be found in the SS433 XRB system.

SS433 was discovered in star surveys to exhibit H$\alpha$ emission in 1977 \citep[see][]{stephenson1977}. It was observed that the emission in this system was Doppler shifted with respect to the position where H$\alpha$ emission is emitted in static material, corrected for Galactic rotation. The emission line for SS433 is threefold: one is shifted to the red, one to the blue, and the third one is not shifted. This is the signature of a static object with two outflows: one oriented toward the observer, and one away from us. This feature appears in almost all objects associated with jets. But what is unique about SS433 is that the shift is evolving with time. Two years after it was discovered, many publications \citep[see][]{fabian1979,milgrom1979a,milgrom1979,margon1979} started investigating the periodic variation of its intensity. The first evidence of periodicity was the 164-day variability. This was explained by the fact that the outflow was changing its position in time, and established the fact that SS433 had precessing bipolar jets. A first estimation of the jet lower velocity was calculated \citep[see][]{fabian1979}, and it was found to be relativistic and higher than $0.2c$.
From emission line observations of SS433, observers refined the velocity for the material of the beam to $0.26c$ and its period to 162 days \citep[see][]{milgrom1982}. These observational evidence led to the kinematic canonical model for SS433: the jet precesses with a period of 162 days with a $20^{\circ}$ precession axis, itself inclined by $78^{\circ}$ from the line of sight (see fig. \ref{geometry}).
Note that this kinematic model adopts a constant relativistic jet propagation speed. In fact, AGN jet simulations often compare propagation speeds with a 1D approximation \citep[see][]{marti1997} based on ram pressure arguments, which distinguishes the jet beam from the jet head advance properties. Furthermore, for a given kinetic luminosity and jet head propagation speed, the density and beam velocity are degenerate with this model. We varied the free parameters for SS4333 to investigate how the jet deceleration varies with them.

SS433 was quickly described as a binary system \citep[see][]{margon1984} composed of a giant star losing its material to its compact companion. The flowing material forms an accretion disk around the smaller partner. This disk gives rise to a collimated jet. The mechanisms by which such jets arise are subject of many studies (see the review by \cite{livio2000,markoff2006}) and are not studied here. Although this picture is widely accepted, the exact nature of the SS433 system has been elusive: evidence would suggest the compact object to be either a neutron star with a low mass or black hole of mass higher than 4.3$M_\odot$  \citep[see][]{margon1984}. More recent studies suggest a mass of up to 16$M_\odot$ (for a total mass of the system of about 40$M_\odot$) \citep[see][]{blundell2008} and favor the black hole scenario. The accretion disk around this black hole is also expected to be of similar size as the orbiting system, namely $10^{12}\mathrm{cm}$ \citep[see][]{krivosheyev2009}.
It is hypothesized that this accretion disk in SS433 is precessing around its compact central object and is unaligned with the orbital plane of the binary. This is discussed in detail in \citet{katz1982}: they matched the shift of the emission lines, corrected for jet contribution, with the prediction from their models. The precession of the disk in turn creates the precessing motion of the jet. An additional nutation of the jet, with a period of 6.5 days, can be linked to the orbital period of the binary, which is known to be 13 days. These details of the jet source region are obtained from observations of the binary system through the analysis of the occultation profile \citep[see][]{margon1979,crampton1981}. We adopted the specific 3D jet morphologies inspired by these SS433 considerations and wish to understand how precession affects the energy transfer from the jet to its surroundings, as opposed to a static jet configuration.

Currently, there exist abundant observations at all scales of SS433. At large scales (20 pc), these reveal the interaction of the SS433 jet with the supernova remnant W50 \citep[see][]{dubner1998}. Hydrodynamic simulations in  3D \citep[see][]{zavala2008} aim to reproduce these observations and argue for a possible time-dependent evolution of the precessing angle for SS433 from 10 to 20$^{\circ}$.
For smaller-scales SS433 observations (about 1 au for the source of the jet), VLBA and VLBI give more insight into the close-in region of the binary. This shows a less continuous jet launch pattern with a high variability as well as X-ray and radio components that are ejected from the central region \citep[see][]{paragi1999}. A movie was released in 2004 showing the time-dependent evolution of its emission over about 40 days (about 1/4 of its period of precession).
For the scales of our study, which are on the order of 0.1pc, we have detailed radio observations. In particular, the observations from the VLA observatory clearly show the helix pattern generated by the precession \citep[see][]{spencer1979,hjellming1981,roberts2008}. Often such observations overplot the kinematic model that is widely adopted for SS433. This model assumes a distance to SS433 of 5.5kpc \citep[see][]{roberts2008}. While this agrees to a large extent with observations, often small discrepancies arise between that model and observational details. We therefore also aim to produce radio maps of our simulations to compare them to observations and the standard kinematic model. 

More recent studies on SS433 were often observational studies on the inner region of the object and circumstellar disk, mainly with the VLBA \citep[see][]{marshall2013,bowler2013}. In contrast, few simulations exist despite the increasing capabilities of modern simulation codes and HPC platforms. \citet{muller2000} performed a $500\times 120\times 120$ resolution, nonrelativistic hydro 3D simulation with a one-sided jet. \citet{zavala2008} performed larger-scale 3D simulation, as mentioned before. Other 2D and 3D simulations focused on the flow of material from the massive star to the compact companion in the binary. Our work aims to investigate VLA observations of intermediate-scale observations of SS433 (about 0.1 parsec) and in future work expand this to the more time-variable smaller scales of current VLBA observations.

In the following, we first introduce our method: the models we used for both hydrodynamic simulations and synthetic radio mapping. This section also presents the tools and diagnostics used to study the simulations. The presentation of our results is divided into three sections: first we givea visual description of the different cases with some first remarks and observations. We continue with the results of a series of diagnostics on our simulations, which provide insight into the energy transfer between the jet and its surroundings and into the role of precession versus variation of the free parameters in our model. The third part of our result presents the radio maps we produced and compares them to actual radio observations of SS433. We finally give some general conclusions of this work.

\section{Simulation aspects}
\label{equation}

As in our previous works \citep{monceau2012,meliani2008}, we used a relativistic hydrodynamic model with the relativistic variant of the Euler equations, with the Mathews approximation \citep[see][]{mathews1976} to the Synge gas relation as a closure. The calculations were performed in a 3D Cartesian grid with a domain spanning $0.2\times0.1\times0.1 \,\,{\mathrm{pc}^3}$ using open boundary conditions on all sides. The bipolar jet was injected time-variably from the center of this domain, using an inner spherical region as a boundary zone, where we fully prescribed all variables within a sphere of radius 0.008 pc at each time. As a result, the source itself is left out of the simulation. This is important for the radio mapping of the simulation, as we show below. We first briefly discuss the numerical approach for the gas dynamical and the radio-mapping part and then explain how we initialized the simulations and analyzed the data in detail.

\subsection{Code and implementation}
\label{code_implementation}

We used the code MPI-AMRVAC\footnote{Code homepage at \textit{\textquoteleft homes.esat.kuleuven.be/\textasciitilde keppens'}} \citep{keppens2011} to numerically integrate the
relativistic HD equations. As explained below, additional equations were solved concurrently for evolving the electron distribution and for passive tracers, which are all easily incorporated into the open-source software\footnote{GIT repository at \textit{gitorious.org/amrvac/amrvac}}.

Our simulations used a TVDLF solver with a threestep time discretization \citep[see][]{toth1996}, where a up to third-order spatial reconstruction is performed with a Koren limiter \citep[see][]{koren1993}. We used a block-based adaptive mesh refinement (AMR) scheme to have full resolution only where needed. We started with a base resolution of $144\times 72\times 72$ and allowed for four levels of refinement, which yielded an effective resolution of $1152\times 576\times 576$ (corresponding in our case to cells of $\Delta x = 1.736 \,10^{-4}$ pc, or eight cells for the jet beam radius at the injection surface). The AMR is triggered automatically by following the evolution of density, pressure and Lorentz factor. Additionally, we always enforced maximal refinement around the central source region of our jet.  
Simulation cases A, B, and C (see below) exploit typically 512 CPUs for 40 hours, case D completed in 20 hours.

\subsection{Model for radio mapping}
\label{radiomodel}

Our method for radio mapping is based on the work of \citet{camus2009} and was implemented in MPI-AMRVAC in \citet{porth2013}. During the simulation we followed the evolution of a passive electron distribution that was injected as a power law $\epsilon^{-\Gamma_e}$ dependence for energies $\epsilon \leq \epsilon_{\infty,0}$, with fixed $\Gamma_e=0.6$.  The initial cutoff energy injected with the jet was $\epsilon_{\infty,0} = 1\rm TeV$.  Three dynamically evolved quantities were added to the simulation, namely $\rho_e(\mathbf{x},t)$, the electron density at time $t$; $\rho_{e0}(\mathbf{x},t)$, the advected initial electron density for the evolved population; and $\epsilon_{\infty}(\mathbf{x},t)$ the cutoff energy. $\rho_{e0}$ simply follows a pure advection equation and was set to initial density values $\rho_{e0}= 1 \mathrm{cm}^{-3}$ in the jet and $10^{-5}\mathrm{cm}^{-3}$ everywhere else. $\rho_e$ follows a conservation law and $\epsilon_{\infty}$ is evolved to account for adiabatic losses. 
Since we did not evolve the magnetic field, the synchrotron losses of $\epsilon_{\infty}$ were neglected in this work. We reproduced radio observations, which are less affected by radiative cooling of the electron population. This cooling mostly affects the high-energy electrons, which determine the X-ray observations. Because the magnetic field is expected to be weak, we also neglected particle accelerations by magnetized shocks and turbulence.

In post-processing we started by reconstructing a new 3D box oriented with respect to the line of sight of the observer. The angle $\theta=78^{\circ}$ between the axis of precession and the line of sight was obtained from radio observations \citep[see][]{margon1984}. This box had a resolution of $64\times 128\times 64$ and was filled by interpolation of the data from the simulation onto the new box. We calculated the synchrotron emissivity in the observer's frame in this new box using equation~(7) from \citet{camus2009},
\begin{equation}
  j_\epsilon = \rho_{e0} D^2 B_\bot (\frac{\rho_{e0}}{\rho_e})^{-\frac{\Gamma_e+2}{3}} \epsilon^{1-\Gamma_e} (1-\frac{\epsilon}{\epsilon_{\infty}})^{\Gamma_e-2},
\end{equation}
where $B_{\bot}$ is the component of the magnetic field normal to the line of sight in the fluid frame, and
\begin{equation}
 D = \frac{\nu_{obs}}{\nu} = \frac{1}{\gamma - \vec{k} \cdot \frac{\vec{v}}{c}}
\end{equation}
is the Doppler factor with $\vec{k}$ the vector of the line of sight, $\gamma$ the local Lorentz factor, $\nu_{obs}$ the observed frequency, and $\nu$ the emitted frequency. The latter is expressed as $\nu = (3 e / 4 \pi m^3 c^5) B_\bot \epsilon^2$, where $e$ and $m$ are the electron charge and mass, and $c$ is the speed of light.
Our simulation does not include the magnetic field, and the 3D geometry of the problem does not suggest a simple magnetic-field topology. Still, we made the approximation to consider a weak frozen-in magnetic field, meaning that in the previous equation $B_\bot$ is taken equal in magnitude to the local pressure value $P$ times a scaling constant of value $10^{-3}$.
After assuming that we have an optically thin medium, we summed this emissivity along the direction of the line of sight. We obtained a 2D intensity radio map for the object with
\begin{equation}
  I(x,y)=\int_{LOS} j_\epsilon(x,y,z) dz.
\end{equation}
To be able to compare our map with observations from a specific telescope, we finally corrected for the instrumental resolution. This was performed by convoluting the emission map with a Gaussian of 0.01pc full width at half maximum (FWHM instrumental convention). This corresponds to the VLA observation restoring beam as used in the observations described in \citet{roberts2008}.

Is it important to note that we only simulated the propagation of the jet and not the central object. As a direct result, the bright central object does not appear on our radio maps. To correct for this, we added a bright pixel at the center of our 2D radio map before applying the beaming. We took the un-resolved flux of that central source to be equal to a fraction of the flux of the obtained map. To empirically match the apparent ratio between the source and the jet intensity as seen in VLA observations, this fraction was taken to be one fourth.

\subsection{Initial conditions}
\label{IC} 

We extracted as many parameters from observation as possible and collected them in Table~\ref{fixparam}. These are identical for all runs and set (1) the thermodynamic conditions of the interstellar medium (ISM) in terms of pressure and density $P_{ISM},\rho_{ISM}$ \citep{safi1997}; (2) the energy flux of the jet $L_{kin}$ \citep{brinkmann2005}, which appears to be typical for an XRB jet; and (3) the jet half-opening angle $\alpha_j$, which we set to $5^\circ$, thus averaging over the fast nutation. Indeed, observations give a half-opening angle for the jet of $1.2^{\circ}$ and $3.5^{\circ}$ for the fast nutation of the jet with period $T_{\mathrm{nut}} \approx 6 \,{\mathrm{days}} \ll T_{\mathrm{prec}} = 162\, \mathrm{days}$. For both jet opening angle and nutation angle see \citet{collins1979,margon1984,borisov1987}.

In \citet{monceau2012} we fixed the density ratio between an injected jet beam and its ISM and derived all other quantities from the given kinetic energy content. Here, we instead fixed the value of the jet Lorentz factor $\gamma_{b}$. Knowing the integrated energy flux over the beam cross section relates to dynamic parameters as
  \begin{equation}
  \label{luminosity}
  L_{kin} = (\gamma_b h_b - 1) \rho_b \gamma_b \pi r^2_b v_{b},
  \end{equation}
where $\rho_b, r_b, v_b$ are the jet beam density, radius and velocity, while $\rho_b h_b = \rho_b + \frac{\Gamma}{\Gamma -1} P_b$ is the enthalpy for a given polytropic index $\Gamma$. $P_b$ was chosen at the injection surface to be equal to the ISM pressure $P_{ISM}$. Knowing $\Gamma$ from the relativistic hydrodynamic equation of state, we can obtain $\rho_b$. This results in a cold supersonic jet.
Table \ref{parameterbinaries} gathers the imposed and derived parameters, which differ from case to case. Here the inertia ratio, defined as $\eta= \gamma_b^2 \frac{\rho_b h_b}{\rho_{ISM} h_{ISM}}$ gives an indication of the difference of conditions between the jet and its surroundings.

\begin{table}[position]
  \begin{center}
  \begin{tabular}{|p{3cm}|p{1.7cm}|p{3.4cm}|}
  \hline 
  \textbf{$L_{kin}$} &  \textbf{$r_b$} & $\rho_{ISM}$\\ 
  \hline 
  $10^{39} \mathrm{erg}.\mathrm{s}^{-1}$ & $0.0007\, \mathrm{pc}$  & $8.3\times 10^{-24} \mathrm{g}.\mathrm{cm}^{-3}$\\ 
  \hline 
  \hline
  $P_{ISM}$& $T_{prec}$& \textbf{Base resolution}\\
  \hline
  $7.5\times 10^{-6} \mathrm{g}.\mathrm{cm}^{-1}.\mathrm{s}^{-2}$ & 162 days & $144\times 72\times 72$\\
  \hline
  \hline 
  \textbf{Domain size (pc)} & $\alpha_j$ &\textbf{Effective resolution}\\ 
  \hline 
  $[-0.1, 0.1] \times [-0.05, 0.05] \times [-0.05, 0.05]$ & $5^{\circ}$ & $1152\times 576\times 576$\\ 
  \hline 
  \end{tabular} 
  \end{center}
    \caption{General parameters for all simulations. See section \ref{IC} for references and a more detailed description.}
      \label{fixparam}
\end{table}

\begin{table}[position]
  \begin{center}
  \begin{tabular}{|p{1.2cm}|p{1.2cm}|p{2cm}|p{1.2cm}|}
  \hline Case & $\gamma_b$ ($v_b$) & Inertia ratio & $\theta_{\mathrm{prec}}$\\ 
  \hline 
  \hline 
  A &  1.036 \, (0.26c) & $28.6$ & 20$^{\circ}$\\ 
  \hline  
  B &  1.87 \, (0.845c) & $0.8$  & 20$^{\circ}$\\ 
  \hline  
  C &  1.036 \, (0.26c) & $28.6$ & 10$^{\circ}$\\ 
  \hline
  D &  1.87 \, (0.845c) & $0.8$  & 0$^{\circ}$\\ 
  \hline
  \end{tabular} 
  \end{center}
    \caption{Parameters for the simulations, which vary from case to case.}
	  \label{parameterbinaries}
\end{table}

\begin{figure}
\FIG{
   \begin{center}
	\centering
      \includegraphics[width=.4\textwidth]{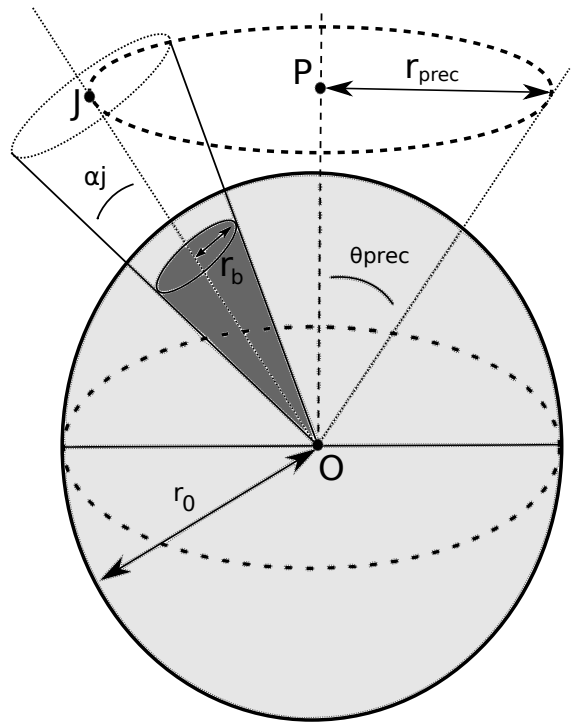}
   \end{center}
}
    \caption{Geometry used for the inner boundary jet-injection prescription.}
	  \label{geometry}
\end{figure}

The injection of the bipolar jet was realized by overwriting at each time step the values within a sphere of radius $r_0 = 0.008 \,\mathrm{pc}$ at the center of our domain, as seen in Figure~\ref{geometry}. We identified the cone of the jet at each time by finding its axis position,
  \begin{equation}
\vec{OJ} = \vec{OP} + \vec{PJ},
  \end{equation}
with $O$ the central source of the jet, $P$ a fixed point on the precession axis, and $J$ a point on the axis of the jet. $\vec{OP}$ is static in time and $J$ has a circular motion around $P$ with a period of 162 days and radius $r_{\mathrm{prec}} = \lVert \vec{OP} \rVert \tan \theta_{\mathrm{prec}}$. After obtaining the instantaneous orientation axis of the jet $\vec{OJ}$, we defined the cone using cylindrical coordinates ($r$,$\theta$,$z$), oriented with the $z$-axis defined by this jet axis. The jet interior obeys  $r \leqslant~ \lvert z \rvert~ \tan~ \alpha_j$. In Figure~\ref{geometry}, the (upper) jet cone is drawn where this cone intersects the sphere of radius $r_0$, and together with the jet opening angle this determines the jet radius on entry $r_b$. We obtained the bipolar jet by simply applying this to negative $z$ as well.
Velocity vectors within the precessing cone are purely radial and the density $\rho$ is defined in the cone of the jet as $\rho(r)=r_0^2 \rho_{b} / r^2$, with $\rho_{b}$ the density as calculated from equation~(\ref{luminosity}). We neglected any rotational component of the velocity, because the rotation speed is much lower than the velocity along the jet axis. The pressure $P$ is defined in the cone as $P(r)=r_0^2 P_{b} / r^2$ with $P_b = P_{ISM}$, following the density variation. The rest of the domain was set to a uniform ISM, as was the part of the central overwritten sphere outside the bipolar jet cone. All parameters for the initial condition and the geometry for all cases can be found in Tables \ref{fixparam} and \ref{parameterbinaries}. 

\subsection{Data analysis}
\label{code_data}

\begin{figure}
\FIG{
   \begin{center}
	\centering
      \includegraphics[width=.45\textwidth]{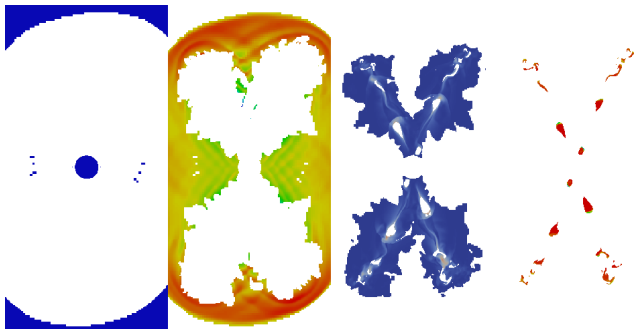}
   \end{center}
}
    \caption{Filter for the region decomposition of the domain. Case A, time $t=1$. From left to right we identify in a cross-sectional plane the unaffected ISM, the shocked ISM, the jet cocoon, and the jet beam.}
	  \label{filter}
\end{figure}

To identify at all times the different regions of the domain affected by jet propagation, we used two complementary methods. A passive tracer followed the evolution of the jet and allowed us to locate purely jet beam material. The second method, previously introduced for our axisymmetric jet simulations in \citet{monceau2012}, uses instantaneous versus initial conditions of the data to locate the entire region altered by jet passage. 
The tracer $\rho_{\mathrm{tr}}$ is passively evolved with respect to the flux of density, hence
\begin{equation}
 \bigtriangledown_\mu (\rho \rho_{\mathrm{tr}} u^\mu)=0\,.
\end{equation}
It was set to the density value in the injection region of the jet ($\rho_{\mathrm{tr0}} = \rho_b$, where $\rho_b$ was obtained from equation \ref{luminosity}) and 0 elsewhere. We can then identify the beam of the jet where the tracer is $\rho_{\mathrm{tr}} \geqslant 0.1 \rho_{\mathrm{tr0}}$. 
This was the filter used to find the instantaneous most remote position of the head of the jet: $r_{\mathrm{head}}(t)$ indicates the farthest radial distance from the central source where jet beam matter exists. Note that for the precessing case, we did not clearly observe the formation of a Mach disk, as found in the study of non-precessing AGN jets or in our case D. Where the value of the tracer is non-zero and $\rho_{\mathrm{tr}} \leqslant 0.1 \rho_{\mathrm{tr0}}$, we found the entire jet cocoon, which also yielded the position of the discontinuity between mixed matter from jet and ISM and the shocked interstellar medium (SISM): hence $r_{\mathrm{disc}}(t)$ is the maximal radial distance where the jet cocoon extends to at a certain time $t$. The shocked ISM is then found from all regions where the energy excess is higher than zero. This energy excess is defined as the difference between the local energy (thermal and kinetic) at the observed time, and the energy at the same location in the initial condition $E_{excess}(\vec{x},t) = E_{tot}(\vec{x},t)-E_{ini}(\vec{x},t=0)$, with $E_{tot} = E_{therm} + E_{kin}$ the total instantaneous energy summing kinetic and thermal energy. Note that this excess energy is the energy measure used below in section \ref{quantification} for energy transfer quantifications. $E_{ini} = E_{therm,ini} + E_{kin,ini}$ is the total energy at time $t=0$ (note that at time $t=0$ the kinetic term is non-zero only in the jet region because the overwritten sphere at the center of the domain is included in the calculation).  This region proivdes the position of the front shock, where $r_{\mathrm{SISM}}(t)$ is once more the farthest radial distance found from this energy excess locator. In locating only shocked ISM region, or only jet cocoon regions, we also ensured that the different regions do not overlap. This implies that we excluded the jet region from all other regions, and similarly excluded the cocoon from the shocked ISM and the shocked ISM from the unaffected ISM region.
An example of a 2D cross section of such filtering for case A at time $t=1$ can be seen in figure \ref{filter}.

\section{Results}

\subsection{General dynamics}

\label{general}

\begin{figure*}
\FIG{
   \begin{center}
	\centering
      \includegraphics[width=.3\textwidth]{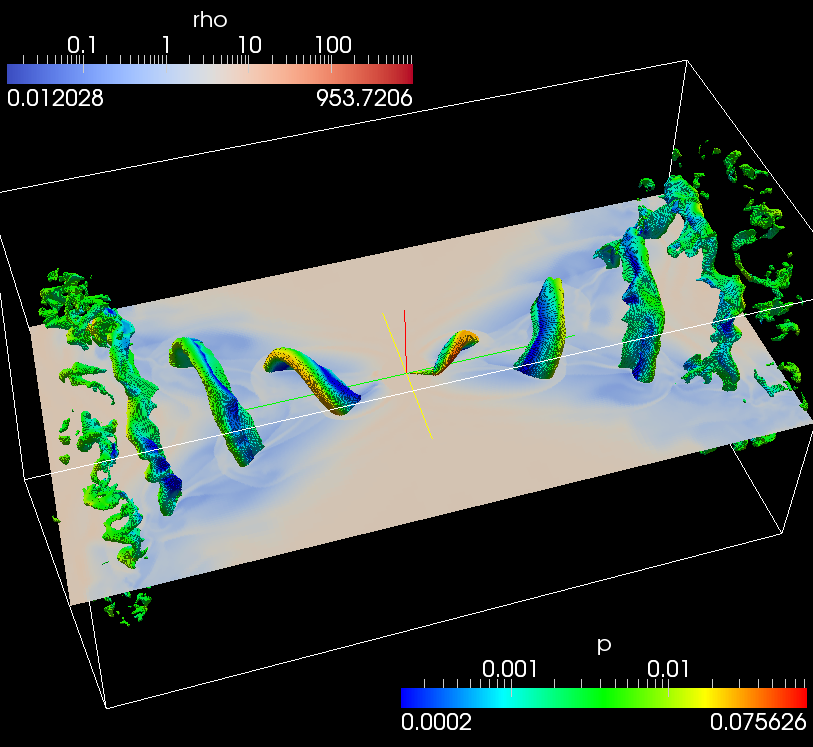}
      \includegraphics[width=.3\textwidth]{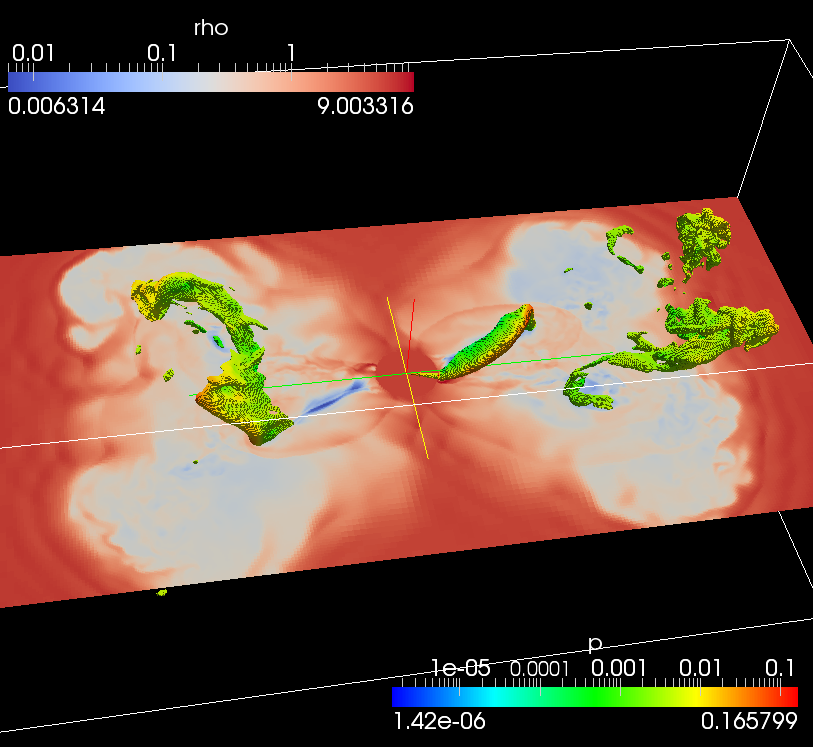}
      \includegraphics[width=.3\textwidth]{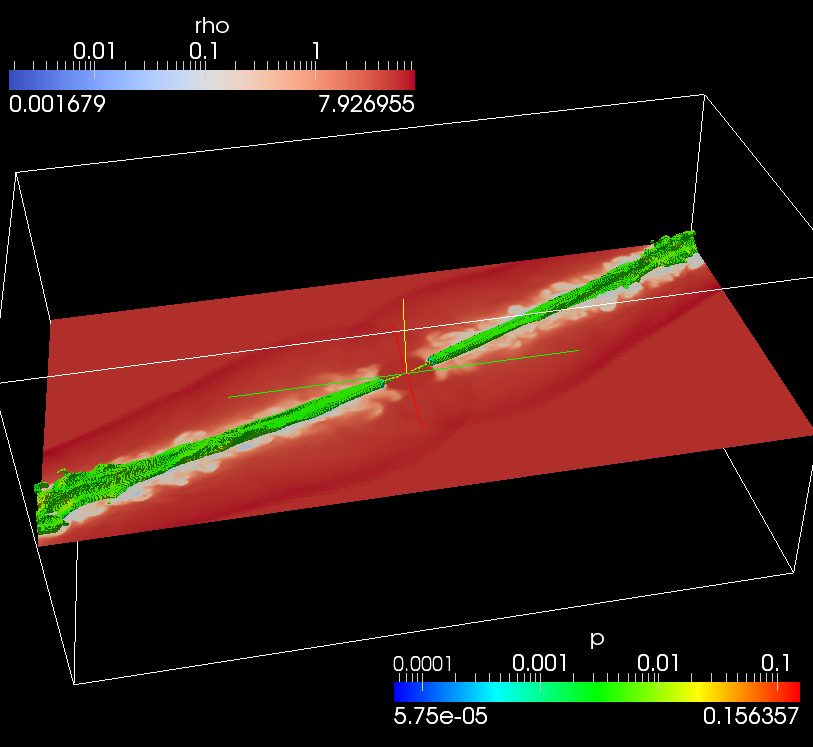}
   \end{center}
}
      \caption{Global view of three cases at times $t=2$ (6.53 years) for case A and B and time $t=0.25$ (0.82 year) for case D. Density is shown on a 2D cut. The jet beam is drawn with the jet filter from section \ref{code_data}. The pressure is quantified on the jet beam surface. Left: case A $\gamma = 1.036$; middle: case B $\gamma = 1.87$; right: case D $\gamma = 1.87$, no precession.}
      \label{global_view}
\end{figure*}

The main feature of the jet associated with SS433 is its precessing motion. This raises the question whether there can be significant differences in dynamics between a static and a precessing jet. Because we considered finite opening angle, and conical jet injections throughout, our previous work, \citet{monceau2012}, provides a description of the expected hydrodynamics of a non-precessing jet, although that work was focused on axisymmetric 2D jets associated with AGN. Here, we considered a jet associated with an X-ray binary. However, it is generally accepted that even if those two cases have completely different scales and overall energetics, their dynamics are similar and the jets are in each case launched from (the central regions of) an accretion disk around a compact object such as a black hole, fed by its surroundings (a galaxy for AGN or a massive companion for XRB).
We recall that for non-precessing, steadily injected relativistic hydro jets, the interaction of the jet with its surroundings includes four important aspects: kinetic energy is transformed into thermal energy through internal shocks in the jet beam (also due to recollimation when injected at finite opening angle) and ahead of the jet at the overarching bow shock. The latter realizes a much more important fraction of the energy conversion. Within the shocked ISM medium, we find many traversing shocks, created at the dynamically evolving contact surface of interaction between the jet and its shocked surroundings, thereby heating the shocked ISM medium. Finally, many fluid instabilities (Kelvin-Helmholtz, Rayleigh-Taylor) develop around the jet, mixing jet and ISM material.
Our first approach to the case of the precessing XRB jet was therefore to directly compare it with a similar, but 3D static one, so that we could see which of these mechanisms were still present, and determine their relative importance.

For the rest of this general discussion, we focus on three of our cases: a precessing jet at the typically quoted SS433 jet speed, case A ($\gamma = 1.036$, $v_b=0.26c$), a precessing case B at higher propagation speed, rendering the jet underdense ($\gamma = 1.87$, $v_b=0.845c$), and its non-precessing analog Case D ($\gamma = 1.87$, $v_b=0.845c$). The comparison between the first two allows one to obtain some insight into the effect of the Lorentz factor on the dynamics of a precessing jet, whereas comparing the last two cases allows for the study of precession itself on the dynamics of mildly relativistic jets. A global representative view on these three cases can be found in figure \ref{global_view}. Cases A and B are visualized at dimensionless time $t=2$ of the simulation, which is 6.53 years in the full evolution, while case D is shown at 0.25, which is 0.82 year, because its jet leaves the domain much earlier. With the period of 162 days for the precession, $t=2$ corresponds to 14.7 precessions of the source. On a 2D plane cut we plot the density, while a volume rendering is realized by using the advected tracer described in section \ref{code_data}. On this volume we quantified pressure.

A first look at the three cases allows a first set of observations: while the non-precessing case D has easily propagated out of the limits of the simulation box already after 0.82 years ($t=0.25$), cases A and B are still fully contained within the domain at $t=2$, with A farther out than B. The general shape of the jet is fundamentally different between the precessing and non-precessing jets: case D has an overall circular cross-section of its beam, well known from extensive studies of relativistic jets of AGN or XRB. The precessing cases, on the other hand, have a helical beam with a ribbon-like appearance with a low-pressure extended part at the back of each annulus. The front side of the annuli display a high-pressure zone. We argue that this pressure pattern is similar to that associated to a propagating piston: high pressure builds in front of the piston, while the pressure drops behind it due to adiabatic expansion. 
Since the size of the simulation box is set to be $0.2\times 0.1\times 0.1 \, \mathrm{pc}^{3}$, with a ballistic propagation where we simply adopted the beam velocity $v_b$, case A and B would cover the 0.1 pc of its maximal expansion in 1.25 and 0.39 year respectively, with 2.8 and 0.9 precessions completed along the way. Both simulations display a higher number of windings through the box, and experienced more precession on both side of their propagation. For case A, we can clearly see four complete windings of precession (648 days) so that the same distance of 0.1 parsec suggests an overall jet speed of propagation of $v_{head}=0.185 c$ (with $c$ the speed of light). This is a clear first hint for deceleration experienced by the precessing jet.

As in the case of AGN jets, one of the main parameters for determining the propagation speed of the jet through the medium is the inertial ratio between the jet and the medium $\eta=\gamma^2 \frac{\rho_b h_b}{\rho_{ISM} h_{ISM}}$, with beam density $\rho_b$ and enthalpy $h_b$, and the ISM density $\rho_{ISM}$ and enthalpy $h_{ISM}$. The ram pressure, which is the force felt by the jet opposing its propagation, is exerted along the axis of propagation and is $F_{ram} \propto - \rho_{ISM} \gamma^2 v_{b}^2$. For a non-precessing jet, it is expected that a higher density ratio results in a more effective deceleration \citep[see][]{marti1997}. For a 1D jet, ram pressure balance at the head allows one to obtain the equation $v_{head}=\frac{\sqrt{\eta}}{\sqrt{\eta}+1} v_b$, where $v_{head}$ is the propagation speed of the head of the jet, $v_b$ is the velocity of the material in the beam of the jet, and $\eta$ is the inertia ratio between the beam, as described above. With a uniform medium and constant beam flux, this results in a steady head velocity. Following that equation for parameters as in case A, a theoretical expected velocity for the jet head is $v_{head}^{marti}=0.22c$, which is higher than the $v_{head}=0.185c$ we previously deduced. 

Recently, \citet{walg2013} improved on the 1D model by taking into account not only the surface interaction of the beam, but the whole surface of the ISM affected by the jet beam, jet head and cocoon included. Therefore, the previous equation was rewritten as $v_{head}=\frac{\sqrt{\eta}}{\sqrt{\eta}+\Omega} v_b$ with the factor $\Omega = \sqrt{\frac{A_{ISM}}{A_j}}$, where $A_{ISM}$ is the surface of the affected ISM and $A_j$ the surface of the jet head. We took this formula to deduce a value of the parameter $\Omega=2$ when the bulk velocity of the jet is 0.185c. Note that the higher density of the jet in case A should indeed not differ too greatly from a bullet-like penetration where $\Omega=1$.

Case D displays a narrow cocoon around its beam with mixing of material between shocked ISM and jet matter. That mixing is also present for the two precessing cases. However, case A shows a somehow similarly narrow cocoon, while case B inflates a wide bubble of low-density material. Only within that cocoon does the jet for case B assume the typical helix shape. The difference between cases B and D, which have the same density ratio, can be explained by their difference in interaction with the ISM. Owing to its precession, case B has a much larger interaction surface than case D. More energy is then transferred to the cocoon for case B, inflating it more than for cases D or A.

The continuous injection regime inside the static-jet case D allows for the formation of standing shocks along the path of the jet. These shocks are the result of recollimation of the jet by the interaction with the surroundings through ram pressure, as already found in axisymmetric studies. These recollimation shocks are virtually absent from precessing jets. As a result, if we still expect ram pressure to play a role for collimation of precessing jets, the inertia contrast will have to be the main way to keep a jet collimated. The inertia of the jet at the injection surface is linked to its density $\rho_b$, itself linked to the Lorentz factor, as seen before. We mentioned that for the same kinetic flux of the jet we can adopt different inertia and Lorentz factor combinations, which is why we included the study of the effect of the variation of $\gamma_b$ on the collimation of the jet.

\subsection{Quantification of dynamics}
\label{quantification}

\begin{figure*}
   \begin{center}
\FIG{
	\centering
      \includegraphics[width=.3\textwidth]{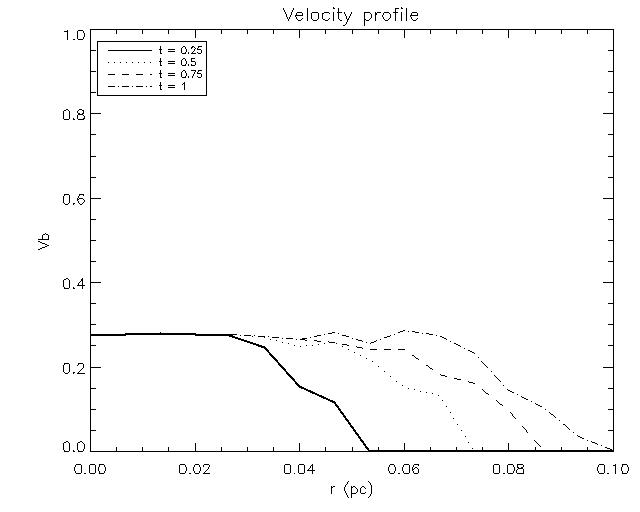}
      \includegraphics[width=.3\textwidth]{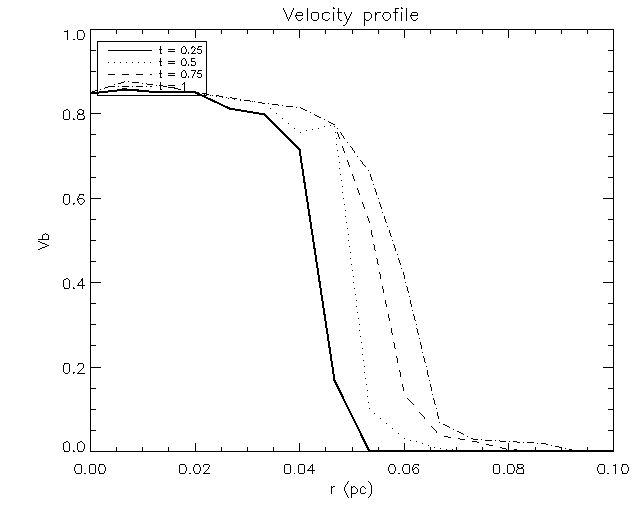}
      \includegraphics[width=.3\textwidth]{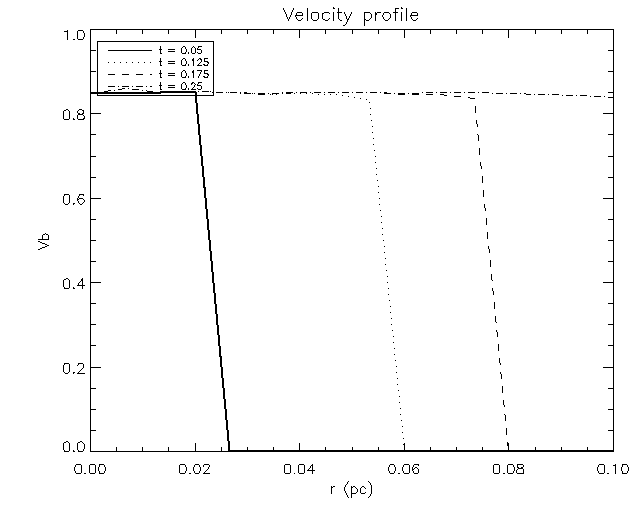}

      \includegraphics[width=.3\textwidth]{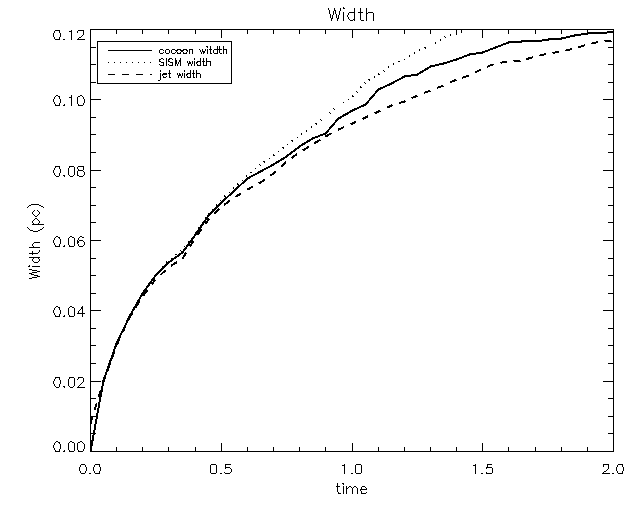}
      \includegraphics[width=.3\textwidth]{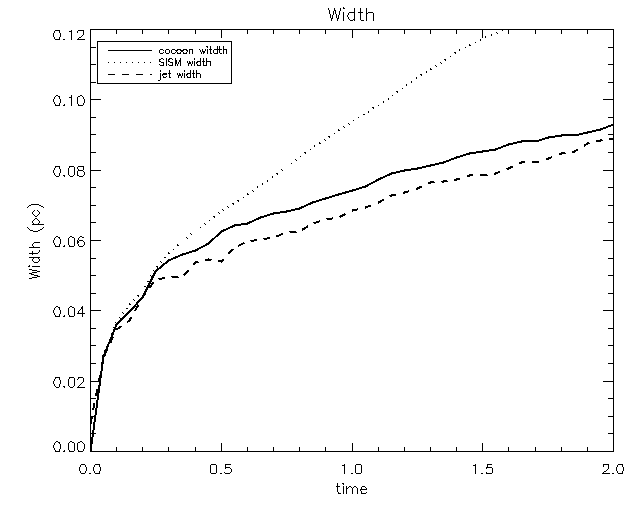}
      \includegraphics[width=.3\textwidth]{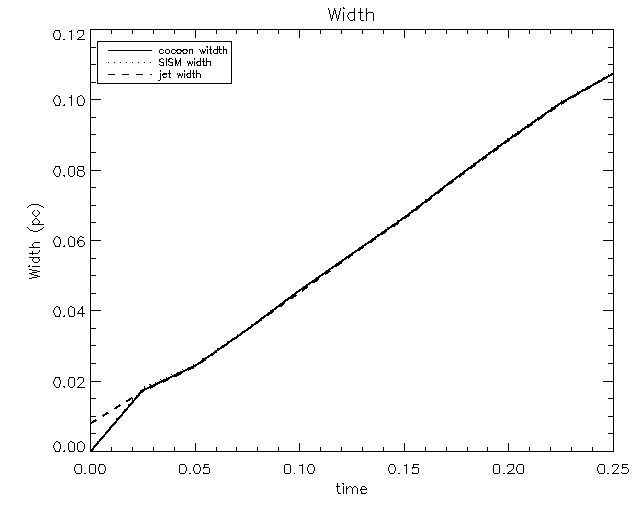}
}
      \caption{Top: Radial speed profiles for various times. Bottom: farthest radial reach of SISM, cocoon, and jet regions. Left: case A $\gamma = 1.036$; middle: case B $\gamma = 1.87$; right: case D $\gamma = 1.87$, no precession.}
      \label{cases}
   \end{center}
\end{figure*}

We now obtain quantified information on the different cases. Figure \ref{cases} gathers the results: in the top row we show the highest velocity profile as a function of radius at four different times for cases A, B, and D. This velocity profile quantifies at each radial distance $r$ from the source $v_b(r)=\max(v(r))$, hence giving the highest velocity value found on a sphere of radius $r$ to the central source point. One of the consequences of its definition is that the two oppositely directed jets both contribute to this measurement. These radial velocity profiles give us information on the internal spatial distribution of the beam velocity $v_b$.
In the bottom-row panels of Figure~\ref{cases}, we use the method described in section \ref{code_data} to show the spatial reach of the jet head, the contact discontinuity, and the front shock as a function of time.

\begin{figure*}
   \begin{center}
\FIG{
	\centering
      \includegraphics[width=.3\textwidth]{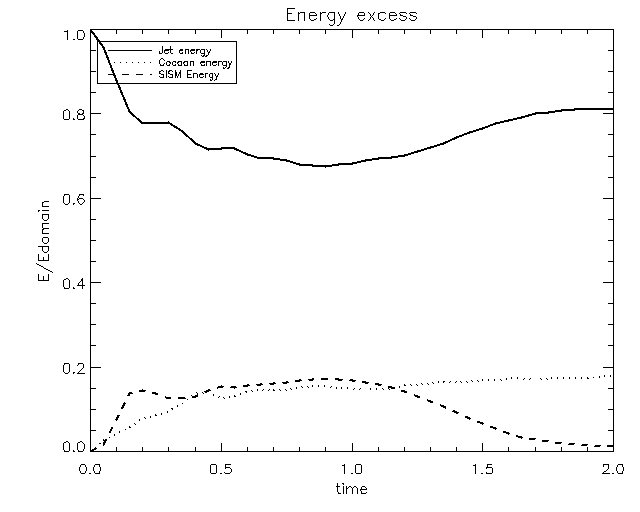}
      \includegraphics[width=.3\textwidth]{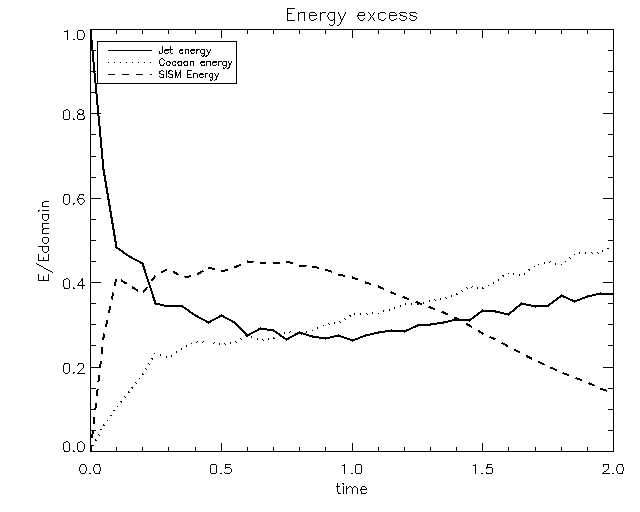}
      \includegraphics[width=.3\textwidth]{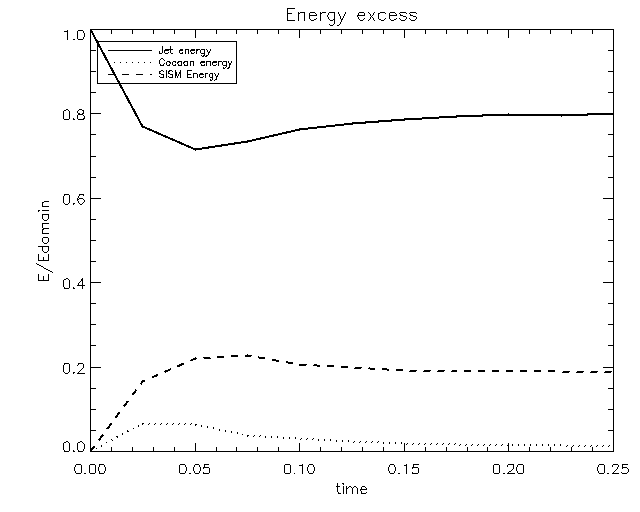}
}
      \caption{Energy transfer evolution as function of time. Left: case A $\gamma = 1.036$; middle: case B $\gamma = 1.87$; right: case D $\gamma = 1.87$, no precession. }
      \label{cases_energy}
   \end{center}
\end{figure*}

The first information given by both the velocity profile evolutions (top row) and radial reaches (bottom row) is on the propagation of the jet head. The bottom row measurements allow us to infer the temporal evolution of the actual head velocity (end limit of the jet) $v_{head}$, the speed of the discontinuity (end limit of the cocoon) $v_{disc}$, and the front shock (end limit of the shocked ISM) $v_{SISM}$, as derivatives of the shown $r_{\mathrm{head}}(t)$, $r_{\mathrm{disc}}(t)$, and $r_{\mathrm{SISM}}(t)$ evolutions.  Note that due to their differences in dynamics, the bottom-row figures are shown up to time $t=0.25$ (0.82 year) for case D and up to time $t=2$ (6.53 years) for both case, A and B.

\begin{figure}[position]
   \begin{center}
\FIG{
      \includegraphics[width=.45\textwidth]{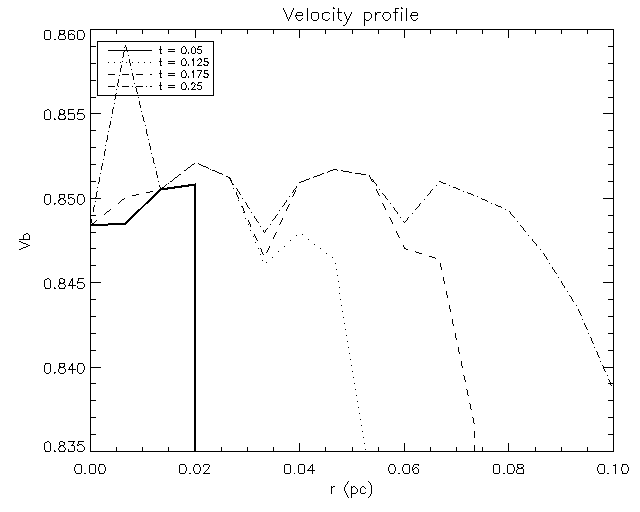}
}
      \caption{Internal structure of the non-precessing case D. Internal structure appears with standing recollimation shocks at 0.02 pc, 0.047 and 0.07 pc.}
      \label{internal_caseD}
   \end{center}
\end{figure}

Case D (no precession, $\gamma=1.87$) shows a very distinct velocity profile throughout: its highest velocity remains nearly constant all along its beam and drops abruptly at the position of its head. A closer look at the velocity profile (Fig. \ref{internal_caseD}) reveals the internal structure of the jet. We can see the formation of the first three standing recollimation shocks at distances 0.02 pc, 0.047, and 0.07 pc.
Its head propagates in accord with the Marti 1D model \citep[see][]{marti1997} prediction with no $\Omega$ correction (see section \ref{general}). Its beam has a velocity of $v_b=0.845c$ and its head propagates with $v_{head}=v_{SISM}=v_{marti}=0.39c$ (as inferred from the lower-right panel of Fig.~\ref{cases}). This case is a canonical underdense relativistic jet, similar to a scaled-down AGN jet. 
While case D follows $v_{marti}$, this is not the case for case B ($v_{marti} = 0.39c$) and case A ($v_{marti} = 0.22c$).
Indeed, case A (left columns, with precession and $\gamma = 1.036$) does not follow the same trend and is decelerated even more as seen from the change in slope of the shown $r_{\mathrm{head}}(t)$. Its radial velocity profile (top-left panel) also shows that its deceleration does not only take place at its head, but all along its beam: closer to the center, its $v_b(r)$ remains unchanged before it smoothly decreases over a large distance. The width of this transition region increases with time: for case A, it measures $0.03 \mathrm{pc}$ at time $t=0.25$ and $0.045 \mathrm{pc}$ at time $t=1$ (3.26 years). We associate this transition region with the SISM region below. 

\begin{figure}[position]
   \begin{center}
\FIG{
      \includegraphics[width=.45\textwidth]{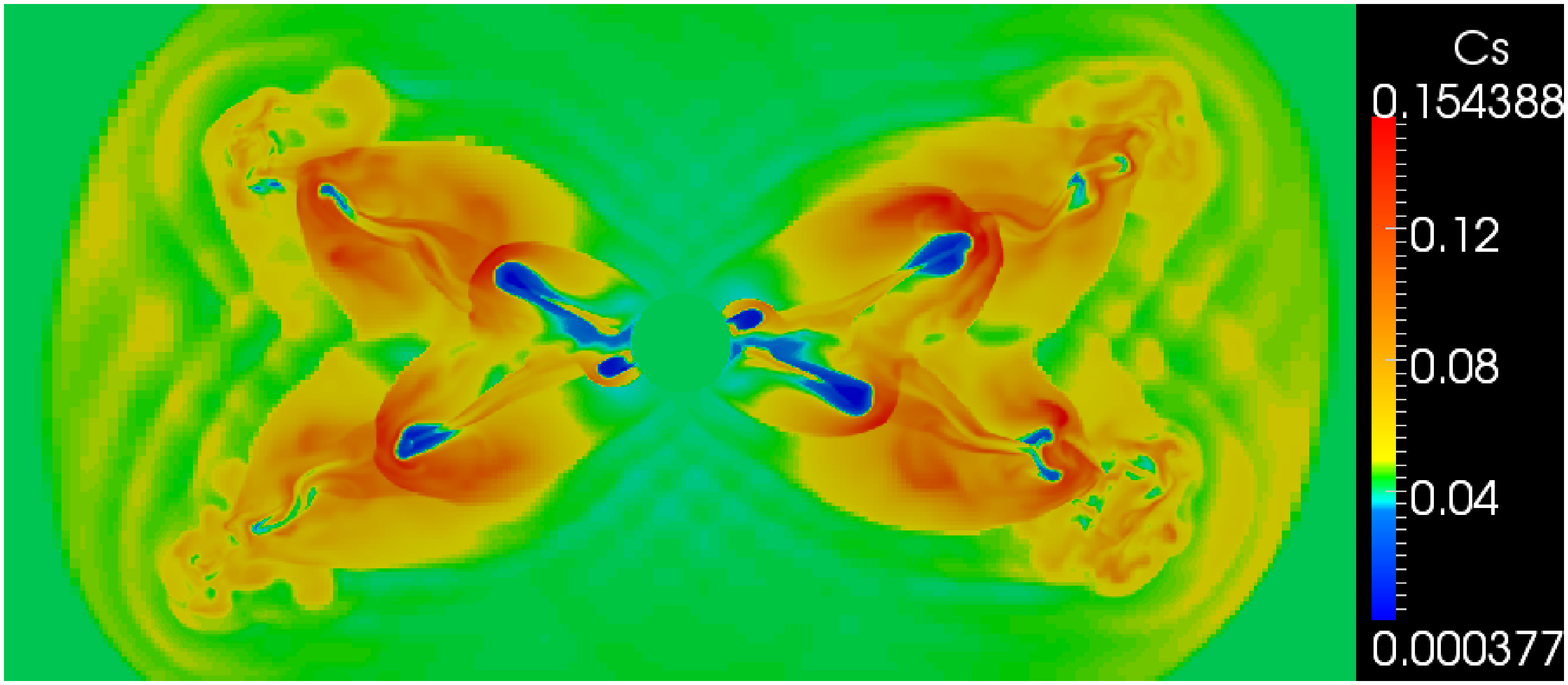}
      \hspace*{0.02cm}
      \includegraphics[width=.45\textwidth]{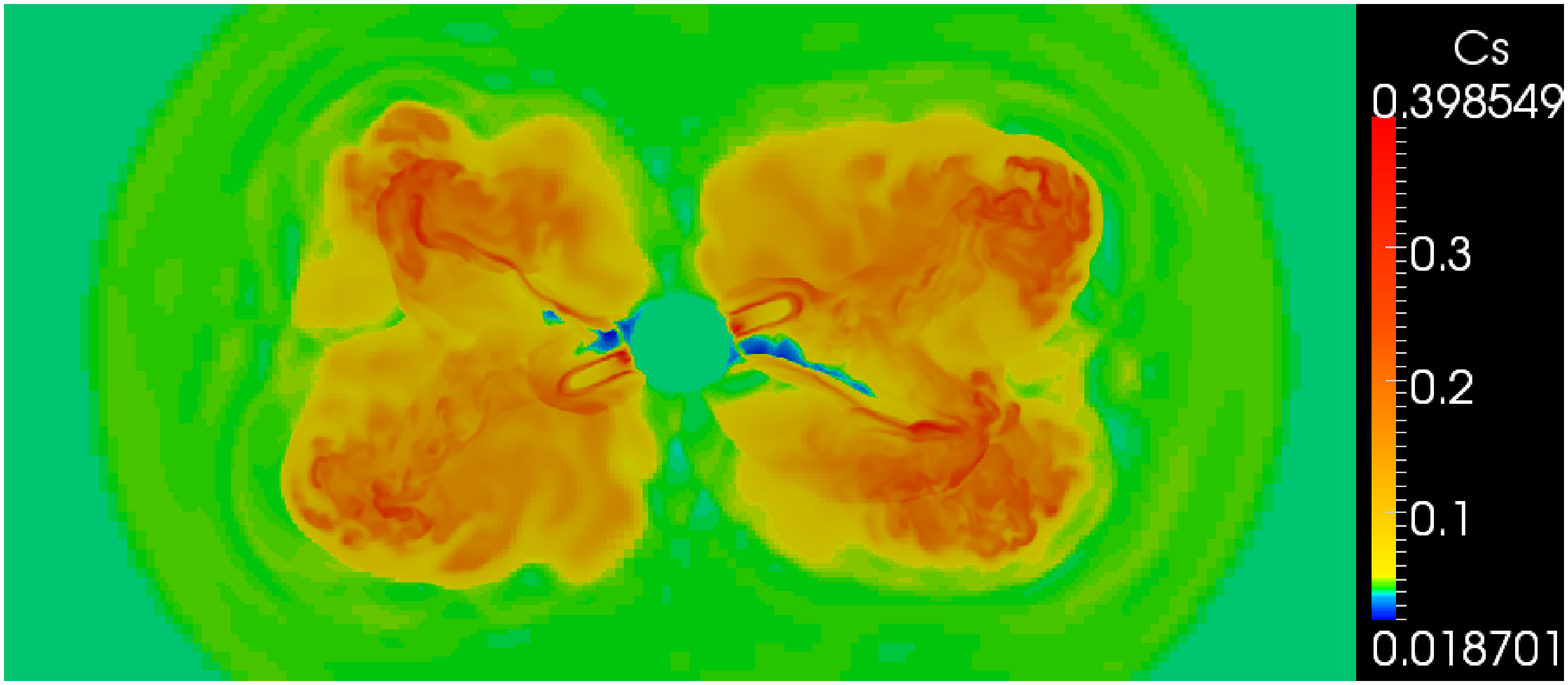}
}
      \caption{2D cut of the domain, mapping of the local sound speed $c_s$ at time $t=1$. Top: case A; bottom: case B.}
      \label{csound}
   \end{center}
\end{figure}

At early times, the jet induces a front (bow) shock and pushes it forward for all cases. This causes in the front shock to initially propagate at the same speed as the jet head supersonic speed. Using a Mach number relative to the ISM sound speed, we measured $M=6$ for case A and $M=22$ for cases B and D. Subsequently, the position of the front shock and the jet head can become disconnected, as the jet head decelerates to subsonic velocity, while the jet beam matter remains supersonic. Case D, as seen before, follows a bullet-like propagation, and we do not observe such a clear disconnection between the three regions during the simulation time. Case A does demonstrate this disconnection at time $t=0.5$ and case B at time $t=0.25$. Then, the velocity of the front shock of these two cases reaches $v_{head}=0.05c$, which is higher than the ISM sound speed $c_{s,ISM}=0.04c$. By plotting the sound speed over a 2D cut of the domain (figure \ref{csound}), we see in cases A and B that the sound speed inside the cocoon is higher than the ISM  sound speed ($c_{s,cocoon} > c_{s,ISM} = 0.04c$). The sound speed in the jet region can exceed $0.1c$, whereas in the SISM region it is on average around the measured velocity for the front shock. The discontinuity between the cocoon and the ISM simply propagates a short distance in front of the transfer surface located by $r_{\mathrm{head}}(t)$ after it disconnected from the front shock. We recall that this transfer surface (or Mach disk in the case of static jets) between the jet matter and the cocoon is the location where conversion from kinetic to thermal energy occurs. Both discontinuity and transfer surface have a similar velocity. This is expected from simple 1D models of relativistic jets. A similar behavior is also present in 2D relativistic jets \citep{monceau2012}. 

Examining the radial velocity profile (top row of fig. \ref{cases}), we can see that it shows a three-part structure: a flat undecelerated profile near the source, a constant deceleration part farther out, and then a third regime where the radial profile displays an ankle as the deceleration softens. This is especially evident for case B: by taking the curve for time $t=1$, one finds the ankle to be at a distance of 0.065 pc and the end of the deceleration zone at 0.09 pc, where the velocity profile reaches zero. The graph of the radial reaches for the same case shows that at time $t=1$ the end of the jet beam and front shock position are found at 0.065 pc and 0.09 pc, respectively. Therefore this region lies outside of the cocoon, in the SISM region. Because the velocity is non-zero, this is direct evidence that the SISM is accelerated to mildly relativistic speed ($0.03c$ for case B). Shocks transfer energy to the SISM region, and we know from previous studies \citep[see][]{monceau2012} that this energy is mainly thermal. But here we see that for precessing jets, they also cause the SISM medium acceleration because they bring kinetic energy to the SISM. The same can be observed for case A. For case D, as already mentioned, the Mach disk and the front shock are not clearly disconnected, thus this feature is absent.

In Figure~\ref{cases_energy}, we used the method explained above to quantify energy deposition aspects.
Two problems arise when considering the energy contents of the different regions for our simulations. Firstly, we used a sharp initial profile in our initial condition from which both SISM and cocoon were ignored, with the jet being directly in contact with the unperturbed ISM at $t=0$. As a result, the simulation needs some time to reach a more physical regime (estimated here for cases A, B, and D to be at times $t=0.15$, $0.15$, and $0.025$, respectively). Secondly, to keep the simulation cost manageable, we used a small box with open boundaries, allowing for dynamics to permeate the continuous boundaries. This allows energy to leave the box at later times of the simulation (estimated here for cases A, B, and D to be at times $t=1$, $1.2$, and $0.23$, respectively). This fact clearly appears in the energy graphs for cases A and B, where the fraction of excess energy present in the SISM-part of the domain decreases rapidly when the SISM begins to permeate the boundaries when both jet and cocoon are still completely inside the domain. These two observations impair our energy measurements at later times. However, we can still reach conclusions from the general trend of the energy graphs. 
The first evident conclusion is reached by comparing cases A and B. In case A most of the injected energy remains in the jet itself, far above both cocoon and SISM regions. In case B, the highest energy fraction is present in the SISM and then in the cocoon, with only one third of it remaining in the jet region. This is a direct consequence of the interaction between the jet and the ISM, as affected by jet precession and its inertia contrast. A much higher fraction of energy is transferred from the jet to the ISM as the Lorentz factor (case A: $\gamma=1.036$; case B: $\gamma=1.87$) increases. This is another means of verifying that case A propagated almost freely in the ISM, whereas case B starts by inflating a hot cocoon in which it can then propagate.
Both cases A and D (with no precession for case D) have a high fraction of energy remaining in the jet.  The two cases have a different inertia ratio between jet and ISM (case A: $\eta=28.6$; case D: $\eta=0.8$). The reason of this low energy transfer is different in origin: for case A we expect the high inertia ratio to cause the jet to travel freely despite the large surface of interaction due to the precession. For case D, on the other hand, the inertia ratio should lead to more interaction. But the surface of interaction is very small compared to all other cases. Without precession we remain with a $\Omega$ factor correction to the 1D Marti equation (see section \ref{general}) equal to 1. But where the cocoon region receives almost no energy in case D, almost 20\% of the deposited energy is transferred to the cocoon for case A. If we sum this with the contribution to the SISM region, we see that about 40\% of the deposited energy is transferred from the jet to the medium for case A, where this is only about 20\% for case D. Again we see that the precession results in an increased interaction between the jet and its surroundings.   

\begin{figure}[position]
   \begin{center}
\FIG{
      \includegraphics[width=.4\textwidth]{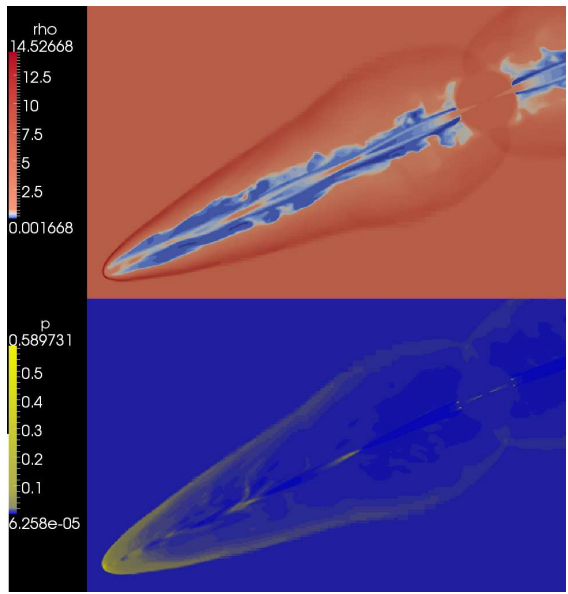}
}
      \caption{Top: density plot of one jet of case D at time $t=0.25$. Bottom: pressure plot of one jet of case D at time $t=0.25$. Formation of a structured beam for the jet by pushing material from the ISM by the head of the jet, as by a piston. Because of adiabatic expansion, the density in that outer layer is low, whereas density and pressure build up in front of the head of the jet. Formation of standing recollimation shocks along the path of the jet.}
      \label{structured_beam}
   \end{center}
\end{figure}

As in \citet{monceau2012}, for the non-precessing case D, we observe (figure \ref{structured_beam}) the formation of a structured beam for the jet with material of the medium being pushed by the head of the jet as by a piston. Because of an extreme adiabatic expansion behind the material of the jet, this outer layer of the beam has low density and pressure. Pressure and density build up in front of the head of the jet. Other known features of relativistic jets can be observed for case D, as seen before: an internal recollimation shock and a strong interaction at the head of the jet. We recall that most of the kinetic energy is transformed into thermal energy at that position. Secondary transfer takes place through the instabilities that appear at the head of the jet and are then advected along it.
 
\subsection{Radio mapping}

\begin{figure*}[position]
   \begin{center}
\FIG{
      \includegraphics[width=.9\textwidth]{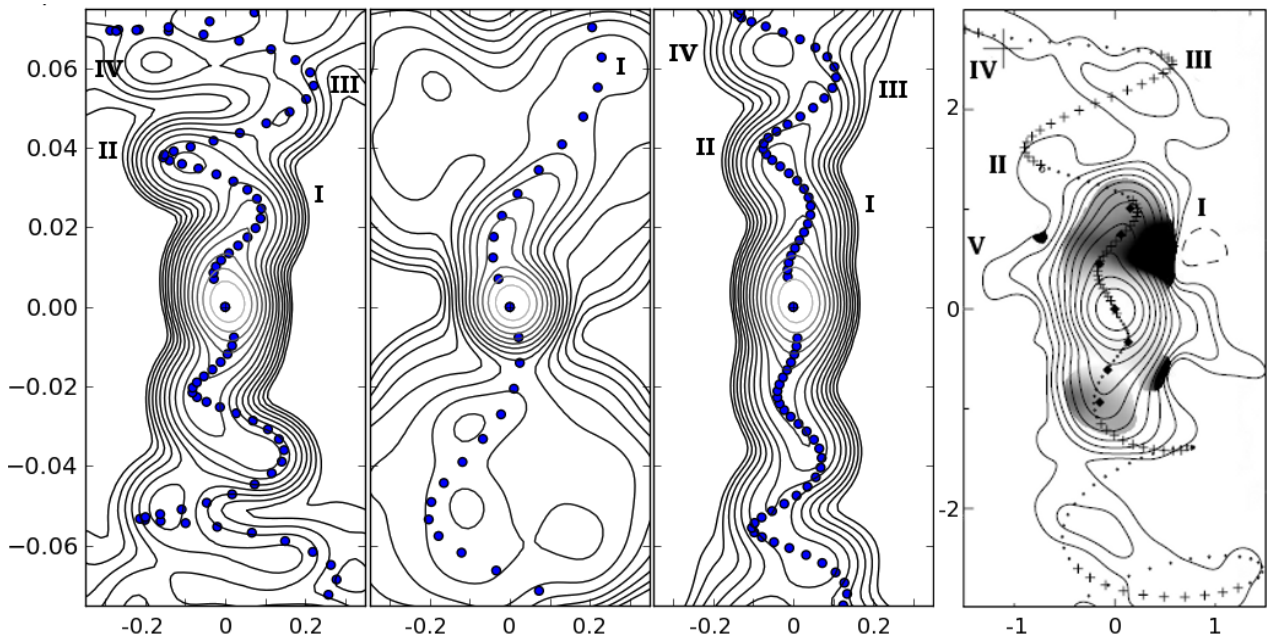}
}
      \caption{Left to right: radio map from simulations cases A, B, and C. Units are given in parsec, the object is estimated to be at a distance of 5.5 kpc. All graphs overplot the kinematic model with parameters corresponding to the case. Right: VLA image of the microquasar SS433 in the constellation Aquila, adapted from \citet{roberts2008}, units are given in arcsecond, with the kinematic model for SS433 overplotted. Both simulated radiomap and observations take contours with steps of factors of $\sqrt{2}$.}
      \label{radiomap-caseABC}
   \end{center}
\end{figure*}

\begin{figure}[position]
   \begin{center}
\FIG{
      \includegraphics[width=.4\textwidth]{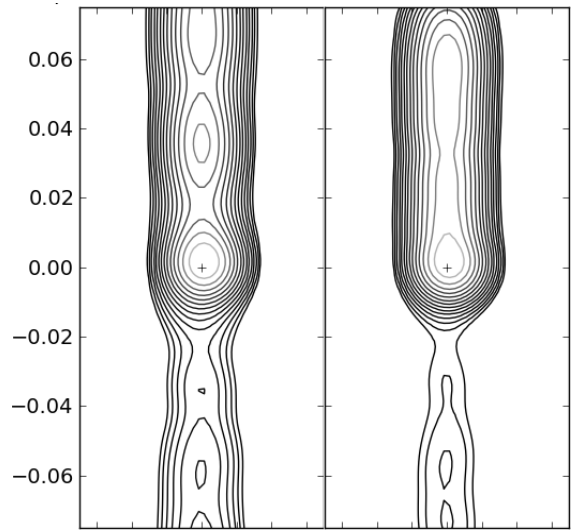}
}
      \caption{Radio maps from the simulation Case D. Because the jet is not related to SS433, we use a different angle to the line of sight to enhance details. Left: angle to the line of sight 85$^{\circ}$. Right: angle to the line of sight 78$^{\circ}$. Distance units are given in parsec. The object is estimated to be at a distance of 5.5 kpc.}
      \label{radiomap-caseD}
   \end{center}
\end{figure}

Following the method described in section \ref{radiomodel}, we realized four different radio maps for cases A, B, C, and D. These radiomaps can be seen in figures \ref{radiomap-caseABC} and \ref{radiomap-caseD}. All cases are construct so that the phase of the precession period is similar. The radio map of cases A, B, and C (which differs from case A only in the precession angle, set to $10^\circ$ for case C) are compared with the observations to test the model. Case D is a good example of the radio emission for a relativistic standing jet.
We know the distance to SS433, which is estimated to be 5.5 kpc. We therefore scaled our radio map to find the correspondence between the projected arcsec and the physical distance in parsec. We note again that for the virtual radio map, we used the pressure as a proxy for the magnetic field. As a result, we can only compare the appearance of the jet features and not their absolute values, such as their intensity. 
In the right part of figure \ref{radiomap-caseABC}, we plot the intensity contours at 4.80 GHz from \citet{roberts2008}, derived from data taken during 1998 March 5-7. Overplotted is the canonical kinematic model assumed for SS433, with plus signs and circles indicating oncoming and receding parts, respectively. We labeled individual bright features that appear in the different figures with I, II, III, and IV. Note that these components, albeit slightly different because of the tilt to the line of sight (78$^{\circ}$ with respect to the axis of precession), are present for both north and south jets. For case B, second from the left in figure \ref{radiomap-caseABC}, only component I is visible. 
Component V, however, does not appear in simulations A, B or C, nor in different observations at different frequencies of SS433. In \citet{roberts2008}, the feature is present at 4.80 Ghz, less extended from 1.415 to 1.665GHz and completely absent at 14.94 Ghz. \citet{doolin2009} studied the hypothesis that this feature might be emitting plasma from the circumbinary disk of SS433. Another hypothesis would be that this feature is linked with backflows at the meeting point of the cocoons inflated by both jets. But we found no evidence for this in any of our simulated cases.

We note that the relative intensities of components I, II, III, and IV are different between the observations and our own radio maps. 
For case A, they are located at the same position, whereas in case B they are located farther away from the source, and in case C they are situated too close to the precession axis (consistent with the smaller precession angle adopted). We note that the locations of these components solely dependent on the dynamics of the jet, while their intensities are strongly dependent on the radiation mechanism. 
In the radio maps, the canonical kinematic model for SS433 is overplotted. We note that the position of the model and the position of the actual bright components are different, even in the observations. This missmatch increases with the distance to the jet source. The reason is that this kinematic model does not take into account the increased interaction between the jet and the ISM due to the precession and the previous deceleration history, which we obtained self-consistently. In contrast to the ballistic behavior of this kinematic model, our simulation case A therefore better matches the observations than the pure canonical model, by taking actual dynamics into account. 
Case B results in a poor match, however, with the radio components escaping the source too fast, as does case C, where the components stay too close to the precession axis. All cases were corrected for Doppler effects. For cases A and C, this is negligible because we see almost no difference in intensity between the upper radio component and their mirror in the lower part of the domain. This is expected because of the low velocity of the jet. By comparison, for case B the radio components in the top part of the domain are much brighter than their counter-part in the lower part of the domain, as the higher Lorentz factor would suggest. If we now compare this with the observation, we see that there is little difference in intensity between the top and bottom radio components for SS433. Both arguments, related to the relative distances and symmetry of the radio components, discard the possibility that there may be a set of parameters for SS433 with a value for $v_{jet}$ different from $0.26c$.
In their model, \citet{zavala2008} used a 10$^{\circ}$ degree precessing angle to reproduce the interaction of SS433 with the supernova remnant W50 at scales of 20pc. The motivation for case C was to see whether such an angle could also reproduce the radio observation at subparsec scale. As shown, this is not the case. We conclude that either there is recollimation of the precessing jet between the two scales, or the angle of precession of SS433 has evolved during its history.


The radio map of case D in figure \ref{radiomap-caseD} shows different features known from the study of relativistic jets: because of the inclination of the jet-axis with respect to the line of sight, the Doppler effect boosts the emission from the jet that points toward the observer and lowers the one that points away from the observer. We can see this pattern for example in the observation of 3C 296 \citep[see][]{laing2006}. 
For statically injected relativistic jets, standing recollimation shocks are known to form along the path of the jet. Case D is no exception, as shown in figure \ref{structured_beam}. These shocks are discussed to be the site of particle acceleration within the jet. This would result in the brightening of these shocks and explain the bright blobs observed along the path of relativistic jets such as in M87 and Cyg A. We recall that we did not account for acceleration of particles in shocks in our simulation (see section \ref{radiomodel}). Nevertheless, we observe in our radio maps of case D (figure \ref{radiomap-caseD}) the appearance of such bright spots. This indicates that the compression of the fluid at recollimation shocks might already suffice for the formation of the emission blobs.

On the other hand, our model for radiation can still be improved; an important step would be to drop the frozen-in assumption and use a passive magnetic field evolved with time. This can be achieved by assuming that it has no retro-action on the dynamics. This would prevent us from moving to a full MHD model and would lower the computational cost.

\section{Conclusion}

We have two separate sets of conclusions for this work. The first conclusions concern fundamental hydrodynamic processes. We showed that the precession of a jet and its Lorentz factor change the interactions with its surroundings and hence the energy exchange between the jet and the medium. The second series of conclusions concern on the observations of the radio maps made from our simulation cases. They showed which set of parameters seems to fit actual VLA observations for SS433 better.

Our hydrodynamic study had two directions: the parametric study evaluated the impact of the variation of the Lorentz factor. We recall that our kinetic flux was constant with the Lorentz factor linked to the density of the jet through equation~(\ref{luminosity}). The first observation is that case A with the lower Lorentz factor ($\gamma = 1.036$) is overdense compared with its surroundings. This resulted in less interaction as the jet propagated easily in the medium. Energy quantification showed that most of the kinetic and thermal energy remained in the jet itself with less than 40\% of the energy in the remaining region of the simulation, SISM and cocoon. By comparison, case B, with the higher Lorentz factor ($\gamma = 1.87$) showed strong interaction. This jet is underdense compared with the ISM, and we observed that more than 70\% of the energy was transferred from the jet to the surroundings. Finally, case D, with its fixed propagation direction, propagated almost freely with less than 30\% of the energy transferred, of which almost nothing entered the cocoon.
As for standing jets in AGN or XRB, the density ratio between the jet and the ISM plays a fundamental role in the interaction between the jet and its surroundings. We recall the three main mechanisms of interaction, or energy transfer, with the first the exchange of kinetic to thermal energy at the front shock of the jet. For non-precessing jets, this mainly takes place at the Mach disk of the jet. The second mechanism is the mixing of material from the jet and the surroundings through instabilities. Finally, shocks formed by the jet-beam propagation can heat the surroundings.
On the other hand, the precession increases these interactions dramatically. We explained this through two aspects: the first mechanism is not present at the jet head alone. We recall that for standing cases it completely dominates the energy transfer \citep[see][]{monceau2012}. Here this particular interaction is present all along the jet. In standing cases, the interaction front can be also larger than the interaction surface of the jet \citep{walg2013}. Here this front is overextended by the precession and the jet interacts with quite a large portion of its surroundings.
For the density ratio, we noted that the helical shape of underdense jets does not survive for as long. This would favor an overdense jet model such as our case A with a Lorentz factor of $\gamma = 1.036$ or an injection velocity $v_j = 0.26c$. It is also interesting to note that we do not know of any quickly precessing AGN jets to be which are widely assumed as being underdense.

Our second set of observations were more directly aimed at reproducing SS433 radio observations. We briefly pointed out that the radio map for standing case D provide some insight into relativistic jets. We showed that the hydrodynamic conditions in the jet alone could already explain some of the features observed: bright radio blobs along the path of the jet could come from pinching of the jets matter at the position of recollimation shocks.
We finally confirmed the scenario of SS433 with a Lorentz factor of $\gamma = 1.036$ or injection velocity $v_j = 0.26c$. This case A visually reproduces the VLA radio observations of SS433 quite well, with bright radio components located at similar distances. On the other hand, we note the importance of taking into account interactions with the medium, which is lacking in the canonical kinematic model. We showed in Fig. \ref{radiomap-caseABC}) that it overestimates the propagation distance of fluid elements in the VLA data. Our model fits the observation better. The reason is that by completely following the full 3D dynamics of the jet as it builds up its entire helical shape, we followed the complex interactions with the surrounding medium, which resulted in a decelerated propagation. This was also found by \citet{panferov2013}, who used an analytic approach to find that to fit the observation, the kinematic model has to be corrected for deceleration. 

In upcoming work we plan to move closer to the source and study possible time-dependent variability of the jet flux. This includes the additional aim to reproduce the time-dependent VLBA observations where blobs are seen to shoot out of the source.

\begin{acknowledgements}
For part of the simulations, we used the infrastructure of the VSC-Flemish Supercomputer Center, funded by the Hercules Foundation and the Flemish Government -department EWI. Part of the simulations used the infrastructure of the Jade - CINES French supercomputer Centre, with a funding from the European project HPC-EUROPA2. Additional simulations were run on Curie and Fermi supercomputers thanks to the PRACE research infrastructure grant SWEET. We acknowledge financial support from FWO-Vlaanderen, grant G.0238.12 and from project GOA/2009/009 (KU Leuven).
\end{acknowledgements}

\bibliographystyle{aa}
\bibliography{article}

\end{document}